# SECUREVENT: Hybrid AI/ML Security Monitoring for Distributed Event-Based Systems

**Eric Liang**
Oracle
zixuan.liang@oracle.com

**Abstract.** Distributed event-based systems have become a common substrate for Internet-scale publish/subscribe services, IoT telemetry, cloud-native microservices, and security operations pipelines. Their loose coupling and asynchronous delivery improve scalability, but they also expand the attack surface: publishers, brokers, subscribers, topics, schemas, and temporal ordering can each be abused without a single component observing the whole behavior. This paper proposes SECUREVENT, a hybrid AI/ML security-monitoring architecture for distributed event-based systems. The architecture combines traditional protections such as authenticated transport, topic-level authorization, and signed events with online anomaly detection, graph-aware behavioral features, complex-event policy rules, federated learning, and adversarial-ML governance. A deterministic prototype study over synthetic event-stream attacks illustrates how a hybrid AI/CEP monitor can improve recall over static rules while retaining a low false-positive rate. The central claim is not that machine learning replaces cryptographic and access-control mechanisms, but that model-based security monitoring is necessary when event flows, identities, schemas, and timing relationships are too dynamic for static controls alone.

## 1. Introduction

Event-based systems decouple producers from consumers by routing messages through topics, subscriptions, content filters, brokers, or event buses rather than direct calls. This design supports dynamic many-to-many communication, but the security boundary becomes more diffuse than in request/response systems. A publisher may not know who consumes an event; a subscriber may not know which producer emitted it; and brokers may need to inspect, route, replicate, buffer, or transform messages while preserving confidentiality and availability. Early work on Internet-scale publish/subscribe security emphasized exactly these tensions: application-level confidentiality, infrastructure integrity, subscription privacy, and denial-of-service resistance are all entangled with the routing fabric itself [1].

Modern event platforms deepen the challenge. MQTT deployments connect resource-constrained devices through lightweight pub/sub protocols; Kafka-style event streaming connects microservices, data pipelines, and security tooling; and complex event processing (CEP) systems correlate events across time. The scale and heterogeneity of these systems make static defenses necessary but incomplete. TLS, authentication, authorization, broker ACLs, signed payloads, schema registries, and encryption reduce known risks, yet they do not reliably detect compromised principals that behave legally but abnormally, colluding publishers that exploit fan-out, replayed events that preserve message syntax, or schema-valid payloads that trigger unsafe downstream workflows.

This paper argues for a hybrid security layer in which AI/ML models analyze event-stream behavior while deterministic CEP and policy rules enforce known invariants. The contribution is fourfold. First, the paper maps security requirements of event-based systems to observable AI/ML feature families spanning identity, topics, timing, payload summaries, graph structure, and broker state. Second, it clarifies how different event-system families create different security signals and failure modes. Third, it proposes an architecture that integrates online anomaly scoring, federated model updates, explainable alerts, and adversarial-ML controls without replacing conventional security mechanisms. Fourth, it reports an illustrative prototype over synthetic event-stream attacks to show how hybrid AI/CEP monitoring can detect attack classes that static rules miss.

## 2. Event-System Taxonomy and Security Implications

A security design for distributed event-based systems should begin by distinguishing the kind of event system being defended. Publish/subscribe brokers emphasize topic membership, publisher and subscriber authorization, and routing confidentiality. Log-centric event streams, exemplified by Kafka's append-only partitioned model [18], emphasize retention, replay, ordering, consumer lag, and backpressure. Stream-processing systems generalize event flow into windowed computations over unbounded data, where correctness depends on event time, processing time, watermarks, and out-of-order handling [19]. Replicated and collaborative systems treat events as state transitions, where causality, clocks, convergence, and conflict resolution become security-relevant properties [20], [21].

These families overlap in practice, but the distinction matters for AI/ML monitoring. A topic-probing attack in MQTT or Kafka may appear as high subscription churn; a replay attack in a log may appear as valid events with abnormal event-time skew; a poisoned collaborative update may appear as a legal operation that changes the causal shape of a document; and a stream-processing abuse case may appear only after aggregation changes a downstream decision. The detector therefore cannot treat every message as an independent packet. It must model identities, topics, schemas, clocks, offsets, windows, replication paths, and consumer effects as parts of one event graph.

| Event family | Security-sensitive state | Typical attack signal | AI/ML monitoring cue |
|---|---|---|---|
| Publish/subscribe | Topic ACLs, routing metadata, publisher/subscriber identities | Topic probing, unauthorized fan-out, spoofed publisher behavior | High-cardinality identity/topic embeddings and churn anomalies |
| Append-only event logs | Offsets, partitions, retention, replay windows, consumer groups | Replay, lag manipulation, partition hot spots, retention abuse | Temporal baselines over offsets, lag, and event-time drift |
| Stream processing and CEP | Windows, watermarks, joins, rule matches, stateful operators | Out-of-order injection, low-and-slow aggregation abuse | Sequence models and rule/model disagreement checks |
| Replicated collaboration | Causal history, conflict resolution, operation logs, merge state | Poisoned operations, collusive edits, causality manipulation | Graph-temporal anomaly detection over operation dependencies |

*Table 1. Event-system families and their security implications.*

## 3. Background and Related Work

Security research on publish/subscribe systems has long identified a split between application security and infrastructure security. Application stakeholders care about who can publish, subscribe, or learn sensitive content; infrastructure operators care about protecting subscription databases, brokers, routing logic, and resource availability [1]. More recent taxonomic work broadens the view beyond pub/sub to event-based systems in general, identifying event spoofing, interception, eavesdropping, confused deputy behavior, collusion, flooding, and delay as event-specific attack families [2]. Confidentiality-preserving pub/sub surveys further show that content-centric routing creates a deep tradeoff: the broker often needs enough information to match events while the application wants to hide content, subscriptions, or both [3].

Operational standards and platform documentation reinforce the need for layered defenses. MQTT 5.0 treats security choices such as network privacy, authentication, and authorization as implementation responsibilities that must be selected for the deployment context [4]. Kafka security documentation likewise centers on SSL/TLS, SASL authentication, and ACL-based authorization [5]. NIST guidance on intrusion detection and prevention frames IDPS technologies as systems that monitor events, identify possible incidents, log evidence, attempt prevention, and report to administrators [6]. These controls are essential, but in event-based systems the relevant evidence is not only packets or host logs; it is the relationship among principals, topics, schemas, ordering, and downstream effects.

AI/ML intrusion detection has already been explored for event-like protocols. MQTT-IoT-IDS2020 showed that flow-level features are important for distinguishing MQTT-specific attacks from benign traffic and from traditional network attacks [7]. Beholder demonstrated that CEP can detect MQTT and CoAP application-layer attacks in IoT settings with precise real-time rules [8]. StreamLearner showed how machine-learning inference can be embedded into distributed CEP for scalable event-stream analysis [9], while SAQL demonstrated that real-time event feeds from enterprise hosts can be queried for rule-based, time-series, invariant, and outlier anomalies at high throughput [10]. The present work builds on this line of research by proposing a general security architecture for event-based systems rather than a detector for one protocol or one dataset.

## 4. Threat Model and Security Objectives

The attacker is assumed to control one or more publishers, subscribers, service credentials, edge devices, or network positions. The attacker may publish syntactically valid messages, subscribe to topics that are permitted by coarse-grained policy, replay old events, induce bursts, delay delivery, probe topic names, poison training data, or adapt behavior after observing alerts. The attacker is not assumed to break modern cryptography; rather, the attacker abuses authorized or partially authorized behavior in a loosely coupled event fabric. This matches a common operational reality: many incidents begin with valid credentials, vulnerable devices, or misconfigured policies rather than cryptographic failure.

The defender seeks five objectives. Authenticity requires that event origins and broker hops be verifiable. Confidentiality requires that payloads, subscriptions, and sensitive metadata be exposed only to authorized parties. Integrity requires protection against event tampering, schema abuse, and unauthorized subscription modification. Availability requires resistance to floods, fan-out abuse, broker exhaustion, and backpressure propagation. Accountability requires durable evidence linking alerts to event IDs, principals, topics, model versions, feature values, and response actions.

| Attack surface | Observable event signals | AI/ML cue | Likely response |
|---|---|---|---|
| Publisher abuse | Unexpected rate, topic novelty, payload drift, schema error bursts | Temporal anomaly and categorical principal/topic embeddings | Throttle, re-authenticate, or quarantine publisher |
| Subscription probing | Rapid subscribe/unsubscribe, denied topics, sparse consumer history | Behavioral sequence deviation and high-cardinality topic encoding | Narrow ACL, require approval, or increase audit logging |
| Broker or routing abuse | Fan-out spikes, unusual broker path, lag, replay, late delivery | Graph anomaly and temporal consistency checks | Reroute, isolate broker, or invalidate replay window |
| Payload and schema abuse | Valid envelope with rare field combinations or abnormal sizes | Schema-aware reconstruction error and robust smoothing | Drop event, route to sandbox, or require consumer confirmation |

*Table 2. Security-relevant event signals and response options.*

## 5. Hybrid AI/ML Security Architecture

Figure 1 summarizes the proposed architecture. The first layer is ordinary platform security: TLS or mutually authenticated transport, identity-bound credentials, topic ACLs, broker authorization, message signatures where end-to-end authenticity is required, and schema enforcement. The second layer is a telemetry tap that exports minimal security metadata from brokers, producers, consumers, and schema registries. The third layer computes stream features and graph features without requiring raw payload centralization. The fourth layer combines AI/ML scoring with CEP rules. The final layer turns high-confidence alerts into response actions and captures analyst feedback for governance.

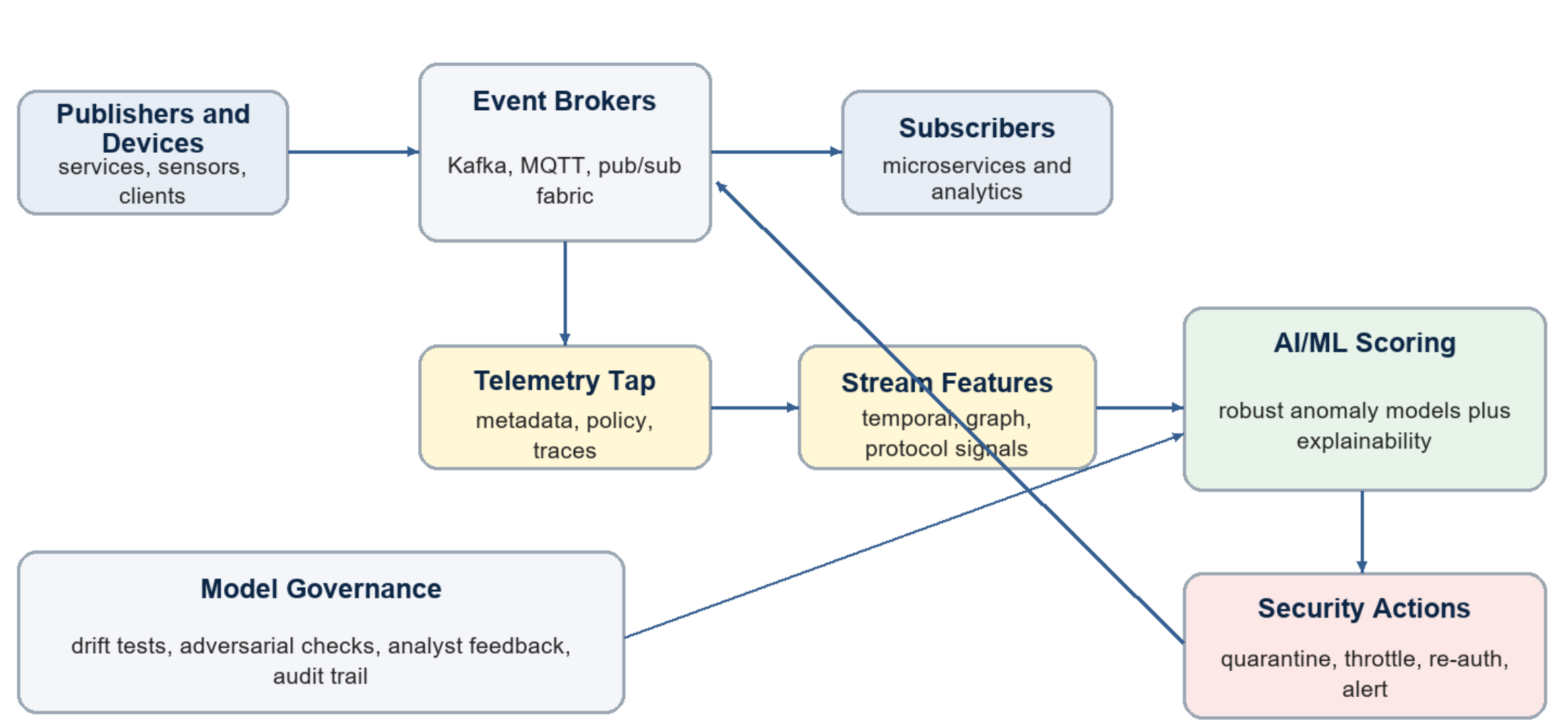

*Figure 1. SECUREVENT hybrid AI/ML and CEP security layer for distributed event-based systems.*

The model layer should not be a monolith. Short-horizon models can detect rate, timing, and lag anomalies; sequence models can capture workflow order; graph models can represent relationships among producers, subscribers, topics, schemas, and brokers; and payload-summary models can detect schema-valid but semantically unusual events. High-cardinality categorical variables are unavoidable because topics, principals, client IDs, device IDs, and schema identifiers often have long-tailed distributions. Efficient categorical representations are therefore central to practical model design, especially when topic and identity spaces evolve rapidly [11].

Metadata governance is equally important. Event models depend on stable definitions of topic names, schema versions, timestamp semantics, principal attributes, and retention policies. Work on harmonizing language-resource metadata is not specific to cybersecurity, but its core lesson applies directly: queryability and accessibility improve when metadata fields are normalized, documented, and mapped across heterogeneous sources [12]. For event security, the same discipline enables consistent feature extraction across brokers, clouds, tenants, and protocol versions.

# 6. Model Design

## 6.1 Feature Families

An event-security model should draw from several feature families. Identity features describe publishers, subscribers, credentials, service accounts, and device posture. Topic and subscription features describe novelty, fan-in, fan-out, hierarchy depth, denied access, and churn. Temporal features describe rate, burstiness, inter-arrival variance, clock skew, late delivery, and replay windows. Payload-summary features describe size, schema version, field presence, entropy-like statistics, and sensitive-field indicators without necessarily inspecting full content. Broker-state features describe partition lag, queue depth, retry behavior, rebalance events, and route changes. Graph features describe how principals, topics, brokers, and consumers relate over time.

Temporal normalization is a surprisingly brittle part of this pipeline. Event sources often mix time zones, string formats, local clocks, Unix timestamps, and broker append times. Automated date format detection for visualization and analytics has direct relevance here because inconsistent time parsing can create false replay alarms, hide actual delays, or corrupt training labels [14]. A production system should preserve original timestamps, normalized timestamps, clock-source metadata, and parser confidence rather than silently overwriting event time.

## 6.2 Detection Strategy

The proposed detector combines three scoring paths. A static policy path captures known violations, such as excessive authentication failures, denied topic probes, schema mismatch bursts, and replay outside a permitted window. A robust unsupervised path learns normal distributions over stream features and reports deviations using median and median absolute deviation statistics, autoencoders, or online density estimators. A relational path computes graph deviations, such as a publisher suddenly reaching many consumers through a rarely used topic. The alert score is a calibrated combination of policy evidence, anomaly magnitude, temporal persistence, and asset criticality.

This hybrid design is intentionally conservative. Static rules alone miss unknown or slow-moving misuse, but pure anomaly detection often produces alerts that operators cannot explain. Combining the two creates a useful division of labor: CEP rules express hard security invariants and known attack signatures, while ML models rank unusual but policy-valid behavior. The alert should expose feature contributions and nearest historical comparisons so that analysts can decide whether to block, rate-limit, route to a sandbox, or simply observe.

### 6.3 Federated and Privacy-Preserving Training

Distributed event-based systems frequently span business units, regions, customers, or administrative domains. Centralizing raw event metadata may violate privacy, contractual, or operational boundaries. Federated learning offers a way to train shared detectors while keeping tenant-specific observations local, following the broader idea of communication-efficient learning from decentralized data [15]. In this paper's architecture, edge domains periodically submit model updates, calibration statistics, or compressed prototypes rather than raw event logs. Secure aggregation and update validation are needed because federated learning itself can be attacked through poisoned client updates.

## 7. Adversarial Robustness and Model Governance

AI/ML expands defensive capability but also adds a new attack surface. NIST's adversarial machine learning taxonomy emphasizes evasion, poisoning, privacy, extraction, and abuse risks across AI life-cycle stages [16]. In event-based systems, poisoning can occur when compromised publishers gradually shift the baseline; evasion can occur when attackers keep each feature just below a threshold; privacy attacks can infer sensitive event distributions from model outputs; and model extraction can reveal which traffic patterns are monitored.

Governance controls should therefore be part of the design rather than an afterthought. Training data must be versioned by time window, tenant, schema, and broker configuration. Model releases should carry audit metadata, validation metrics, drift tests, and rollback plans. Analyst feedback should be treated as labeled evidence, not as immediate retraining input. Certified robustness techniques, including improved certified-radius estimation for randomized smoothing, provide one direction for quantifying how much perturbation a classifier can tolerate before its prediction is no longer certified [13]. Such certification is not a complete defense for event-stream security, but it is a useful principle: model deployments should expose measurable robustness claims rather than only aggregate accuracy.

Explainability is also a safety requirement. A security team cannot responsibly quarantine an event producer if the alert says only 'anomaly score 0.93.' Local explanations, counterfactual examples, feature contribution summaries, and lineage links to specific events make alerts auditable. This echoes the broader explainable-ML argument that users need reasons for individual predictions, not only global model performance [17].

## 8. Prototype Study

To make the architecture concrete, a deterministic prototype generated 3,600 five-second event windows containing benign traffic and six injected attack episodes: event flood, authentication brute force, topic spoofing, exfiltration-like fan-out, replay/delay, and collusion. Each window contained nine aggregate features: publish rate, subscribe rate, topic novelty, authentication failures, schema errors, mean payload size, fan-out, late-event ratio, and principal churn. The first 500 benign windows trained a robust profile using medians and median absolute deviations. The study then compared static CEP rules, a streaming robust-ML detector, and a hybrid AI/CEP monitor.

| Method | Precision | Recall | F1 | False positive rate | Median latency (s) |
|---|---|---|---|---|---|
| Static CEP rules | 1.000 | 0.528 | 0.691 | 0.00% | 0 |
| Streaming robust ML | 0.919 | 1.000 | 0.958 | 1.21% | 0 |
| Hybrid AI/CEP monitor | 1.000 | 0.973 | 0.986 | 0.00% | 0 |

*Table 3. Prototype detection metrics on synthetic event-stream attacks.*

**Prototype Detection Performance on Synthetic Event-Stream Attacks**

False positive rates: Static CEP rules 0.00%; Streaming robust ML 1.21%; Hybrid AI/CEP monitor 0.00%

*Figure 2. Precision, recall, and F1 for static rules, robust ML, and hybrid monitoring.*

The results are not intended as a benchmark claim because the workload is synthetic. They do, however, illustrate an important design pattern. Static rules achieved high precision but missed some policy-valid anomalies, especially exfiltration-like fan-out and collusion. The robust ML detector improved recall but introduced more false positives. The hybrid monitor preserved most of the precision of rules while recovering much of the recall of ML. This is the practical value of hybridization: the system can use deterministic rules where security semantics are known and AI/ML where distributed behavior is too varied to enumerate.

## 9. Deployment Considerations

Deployment should begin with observability, not automatic blocking. A useful first stage is shadow mode, where models score events and generate explanations without changing routing. Operators can then compare alerts against incident tickets, broker metrics, and service-owner feedback. After calibration, response actions should be tiered: low-confidence anomalies create enriched alerts; medium-confidence anomalies increase logging or require re-authentication; high-confidence anomalies trigger throttling, quarantine topics, or broker-side circuit breakers. Automatic deletion of events should be rare because the event stream itself is often forensic evidence.

The architecture should also account for latency budgets. Inline inference may be suitable for critical topics with low throughput, but high-volume telemetry often requires asynchronous scoring or broker-adjacent sampling. Model state must be partition-aware so that rebalances and failovers do not erase context. Feature stores must preserve both event time and processing time. Finally, security teams should test detection under broker failure, consumer lag, schema evolution, clock skew, and adversarially shaped low-and-slow attacks, not only under obvious floods.

## 10. Limitations and Research Agenda

Several open problems remain. First, public datasets rarely capture realistic multi-tenant event systems with broker state, schema changes, authorization decisions, and downstream effects. Second, labels are sparse because real incidents are rare, ambiguous, and often investigated outside the event platform. Third, concept drift is structural: new services, topics, schemas, and consumer groups appear continuously. Fourth, encrypted payloads restrict feature extraction, requiring a careful balance between confidentiality and detection. Fifth, federated detection raises governance questions about poisoned updates, tenant weighting, and cross-domain comparability.

Future research should therefore focus on realistic event-security testbeds, privacy-preserving feature extraction, graph-temporal foundation models for event streams, certified robustness for stream classifiers, and operator-centered alert explanation. Another promising direction is to integrate policy-as-code with model scoring so that every automated response can be traced to both a policy condition and a model explanation.

## 11. Conclusion

Distributed event-based systems need security mechanisms that match their architecture. Traditional controls authenticate principals, encrypt channels, and authorize topics, but they cannot fully observe dynamic misuse across asynchronous, many-to-many flows. AI/ML models can fill this gap by learning behavioral baselines, detecting graph and temporal anomalies, and prioritizing suspicious events for analyst review. The safest design is hybrid: deterministic CEP rules enforce known security invariants, while AI/ML models detect policy-valid behavior that is operationally abnormal. When combined with metadata governance, federated training, adversarial robustness, and explainable response actions, this approach can improve the security posture of modern event fabrics without pretending that machine learning is a substitute for sound security engineering.